\documentclass{article}

\usepackage{arxiv}

\usepackage[utf8]{inputenc} 
\usepackage[T1]{fontenc}    
\usepackage{hyperref}       
\usepackage{url}            
\usepackage{booktabs}       
\usepackage{amsfonts}       
\usepackage{nicefrac}       
\usepackage{microtype}      
\usepackage{lipsum}     
\usepackage{graphicx}
\usepackage{doi}

\usepackage{amsmath}
\usepackage{amssymb}
\usepackage{latexsym}
\usepackage{graphicx}
\usepackage{natbib}
\usepackage{color}
\bibpunct{(}{)}{;}{a}{}{,}

\usepackage{amstext}

\usepackage{adjustbox}

\usepackage{amstext}

\newcommand{\arcsec}{\ensuremath{^{\prime\prime}}}
\newcommand{\apj}{ApJ}
\newcommand{\aap}{A\&A}
\newcommand{\apjl}{ApJL}
\newcommand{\apjs}{ApJS}
\newcommand{\nat}{Nature}
\newcommand{\solphys}{Sol. Phys.}

\newcommand{\jgr}{J. Geophys. Res.}
\newcommand{\mnras}{MNRAS}

\title{Estimating the Poynting flux of Alfv\'enic waves in polar coronal holes across Solar Cycle 24}

\author{ \href{https://orcid.org/0000-0001-5678-9002}{\includegraphics[scale=0.06]{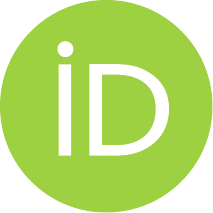}\hspace{1mm}Richard J. Morton}\\
Department of Maths Physics and Electrical Engineering \\
Northumbria University, UK \\
\texttt{richard.morton@northumbria.ac.uk}\\
\And
\href{https://orcid.org/0000-0002-4433-4841}{\includegraphics[scale=0.06]{orcid.pdf}\hspace{1mm}M. J. Weberg}\\
US Navel Research Laboratory, Washington DC, 20375, USA \\
\And
N. Balodhi\\
Department of Mathematics, Physics and Electrical Engineering, Northumbria University, UK\\
\And
\href{https://orcid.org/0000-0002-7863-624X}{\includegraphics[scale=0.06]{orcid.pdf}\hspace{1mm}J. A. McLaughlin}\\
Department of Mathematics, Physics and Electrical Engineering, Northumbria University, UK \\
}

\begin{document}
\maketitle

\begin{abstract}
Alfv\'enic waves are known to be prevalent throughout the corona and solar wind. Determining the Poynting flux supplied by the waves is required for constraining their role in plasma heating and acceleration, as well as providing a constraint for Alfv\'en wave driven models that aim to predict coronal and solar wind properties. Previous studies of the Alfv\'enic waves in polar coronal holes have been able to provide a measure of energy flux for arbitrary case studies. Here we build upon previous work and take a more systematic approach, examining if there is evidence for any variation in vertical Poynting flux over the course of the solar cycle. We use imaging data from SDO/AIA to measure the displacements of the fine-scale structure present in coronal holes. It is found that the measure for vertical Poynting flux is broadly similar over the solar cycle, implying a consistent contribution from waves to the energy budget of the solar wind. There is variation in energy flux across the measurements (around 30\%), but this is suggested to be due to differences in the individual coronal holes rather than a feature of the solar cycle. Our direct estimates are in agreement with recent studies by \cite{Huang_2023,Huang2024} who constrain the vertical Poynting flux through comparison of predicted wind properties from Alfv\'enic wave driven turbulence models to those observed with OMNI at 1~AU. Taken together, both sets of results points towards a lack of correlation between the coronal Poynting flux from waves and the solar cycle.
\end{abstract}



\section{Introduction}\label{sec:intro}

Alfv\'enic waves are thought to be a key mechanism for the transport of energy through the solar atmosphere, supplying energy for heating the coronal plasma and the acceleration of the solar wind \citep{Van_Doorsselaere_2020b,Morton_2023}. Over the last decade or so it has become well established that propagating Alfv\'enic waves\footnote{Throughout this work the term Alfv\'enic is used to discuss the transverse motions observed in the corona. The observed waves have been interpreted in terms of the MHD kink mode which occurs in a structured plasma \citep{SPR1982,EDWROB1983}. However, this mode is Alfv\'enic in nature and is effectively a surface Alfv\'en mode \citep{GOOetal2012}.} are present throughout the solar atmosphere \citep{TOMetal2007,MCIetal2011,MORetal2019}. Numerous studies indicate Alfv\'enic modes appear consistently in coronal holes \citep[e.g.,][]{THUetal2014,MORetal2015,Weberg_2020}, which are known to be the source regions for fast wind streams. The crucial role of Alf\'venic waves is supported by numerous numerical models that have been able to demonstrate Alfv\'en wave turbulence is capable of producing hot, fast winds \citep[e.g.,][]{SUZINU2005,CRAetal2007,Chandran_2011,van_Ballegooijen_2016,Shoda_2019}. Moreover, there are several Alfv\'en wave turbulence driven magnetohydrodynamic (MHD) models which are able to broadly reproduce observed wind properties when using magnetograms and magnetic maps to define the inner boundary conditions \citep[for example, the Alfvén Wave Solar Atmosphere Model - AWSoM,][]{van_der_Holst_2014,Sachdeva2019}. Additionally, the Alfv\'en wave driven wind models are increasingly used for predictive purposes, such as estimating solar/stellar evolution \citep[through mass and angular momentum loss, e.g.,][]{Hazra_2021,Shoda_2023}, planetary habitability around stars \citep[e.g.,][]{Garraffo_2016,Shoda_2021}, and for forecasting space weather conditions \citep[e.g.,][]{2023SpWea..2103262J}. 

\medskip

It is known that the properties of the solar wind depend upon the level and location of energy deposition \citep{LEEHOL1980}. \textbf{In principle, one needs to obtain the full Poynting flux vector to understand the total energy input into the solar wind. This is a challenging prospect in the low corona as it requires knowledge of the vector and magnetic field vectors. However, for Alfv\'enic waves the group velocity is directed only along the magnetic field \citep[e.g.,][]{GOEPOE2004} hence we only need to estimate the vertical component of the Poynting flux \citep[e.g.,][]{GOOetal2013}.} 

 For Alfv\'en wave driven models, the solar wind speed, density and levels of mass and angular momentum loss are directly related to the vertical Poynting flux input at the coronal base \citep[e.g.,][]{Hazra_2021}. In these models, the vertical Poynting flux ($\rho v_A \delta v^2$) has two free components, the magnetic field strength (via $v_A=B/\sqrt{\mu_0\rho}$) and wave amplitude ($\delta v^2$) in the corona. While the magnetic field strength can be estimated reliably from magnetograms (e.g., via potential field extrapolations), the Alfv\'enic wave amplitude is still a free parameter. \cite{Hazra_2021} took the approach of holding the wave amplitude constant, which led to solar cycle dependent mass and angular momentum losses. This was because the Poynting flux varied due to the increase and decrease in surface magnetic field. Another study by \cite{Huang_2023,Huang2024} set about estimating the energy flux required for AWSoM to match solar wind observations recorded with OMNI at 1~AU. The work used photospheric synoptic magnetic maps from different time periods across the solar cycle \citep[as did][]{Hazra_2021}, varying the vertical Poynting flux injected into the model to best match solar wind conditions (the level of vertical Poynting flux in each model run was controlled by their so-called Poynting flux parameter, which is equivalent to changing the coronal wave amplitude). The results suggest that an energy input of $\sim500$~W~m$^{-2}$ is required at the base of the corona\footnote{Note the original values in \cite{Huang_2023,Huang2024} are a factor of 10 to small, which has been confirmed with the authors.}. In contrast to \cite{Hazra_2021}, the required coronal wave amplitude had to be varied between data sets in order to match observations.

\medskip

Given the different approaches, it is important to determine if the properties of Alfv\'enic waves at the Sun are varying over time and show any particular patterns, constraining Alfv\'en wave turbulence models and improving the reliability of predictions about solar wind conditions. While the previous observational studies of the coronal Alfv\'enic waves have probed the presence, amplitudes, frequencies and energy flux of the waves associated with specific time periods, it is unknown whether there is any variation in wave energy flux into the coronal holes over the solar cycle. Results from \cite{MORetal2019} using the Coronal Multi-channel Polarimeter (CoMP) indicate that Alfv\'enic waves are present globally throughout the solar cycle, with some variation in the global properties of the waves' Fourier power spectra. However, information on waves in coronal holes was largely missing from that work. This was due to a combination of the temperature sensitivity of the Fe XIII infra-red line that is observed by CoMP ($\sim1.6$~MK in ionisation equilibrium) and the instrument sensitivity. Coronal hole plasma is generally relatively cool \citep[$\sim 1$~MK below 1.3~$R_\odot$, e.g.,][]{Landi_2008,Saqri_2020}, so only has weak emission for lines formed at higher temperatures below 1.3~$R_\odot$.

\medskip

The wave energy flux entering into the coronal holes will likely be determined by the driving mechanism and/or plasma and density structure in the lower solar atmosphere. One of the main candidates for exciting transverse waves is the buffeting of magnetic field lines by the convective photospheric flows. The magnetic funnels that define the structure of coronal holes are believed to originate from the 
network magnetic fields \citep{Hassler1999,Tuetal2005}. The network is thought to be supplied by the emergence and advection of internetwork fields by the supergranular flows \citep{GOSetal2014}. The origin of the internetwork fields is still debated but there is evidence that their properties remain consistent over the solar cycle, hinting they are produced by a local, sub-surface dynamo \citep{Buehler2013,Faurobert2015}. To date there appear to be only a few investigations into cycle variations of the dynamics of the photosphere and convection, with contrasting results. \cite{Roudier_2017} suggest that the properties of meso and supergranule scales appear near constant over the solar cycle.  \cite{Muller2018} find little change in granulation contrast and scale over the solar cycle, with \cite{Ballot2021} finding variations in density and area are of the order of $\sim2\%$. However, \cite{Getling_2022} report that the subsurface convective flow displays variations in the integrated power of the velocity field over the solar cycle.

Another potential driver for coronal Alfv\'enic waves is \textit{p}-modes, which are thought to undergo mode conversion in the lower solar atmosphere \citep{CALGOO2008,KHOCAL2012,CAL2017}. Estimates of the Fourier power spectra of the coronal Alfv\'enic waves from CoMP show a clear enhancement of power at $\sim4$~mHz \citep{TOMetal2007,MORetal2019} which could be an indicator of this process. Although a clear link between the observational signal and the mode conversion process is still missing. There have been a number of studies into the variability of \textit{p}-mode properties over the solar cycle. The \textit{p}-modes are thought to be generated stochastically and damped by the near surface turbulent convection. Variation in magnetic fields over the solar cycle can influence the convection and hence influence the behaviour of \textit{p}-modes. This is found to be the case for sun-as-star measurements of the \textit{p}-mode parameters. Variations in mode frequency, width and amplitude over the solar cycle are now well established \citep[e.g., ][]{Howe_1999,Komm_2000,Kiefer2018}. In particular the mode energies and mean squared velocities are anti-correlated with solar activity (e.g., the 10.7~cm radio flux) and vary by 18\% and 14\% respectively \citep{Kiefer2018}. However, as mentioned, these measurements are performed on sun-as-star data, and include the influence of active regions, plages, etc. Hence the observed variations in \textit{p}-mode properties from such studies might not reflect the local properties of the modes in quiet Sun and coronal hole regions. 

\begin{figure}[!t]
    \centering
    \includegraphics[trim=20 0 5 30, clip, scale=0.5]{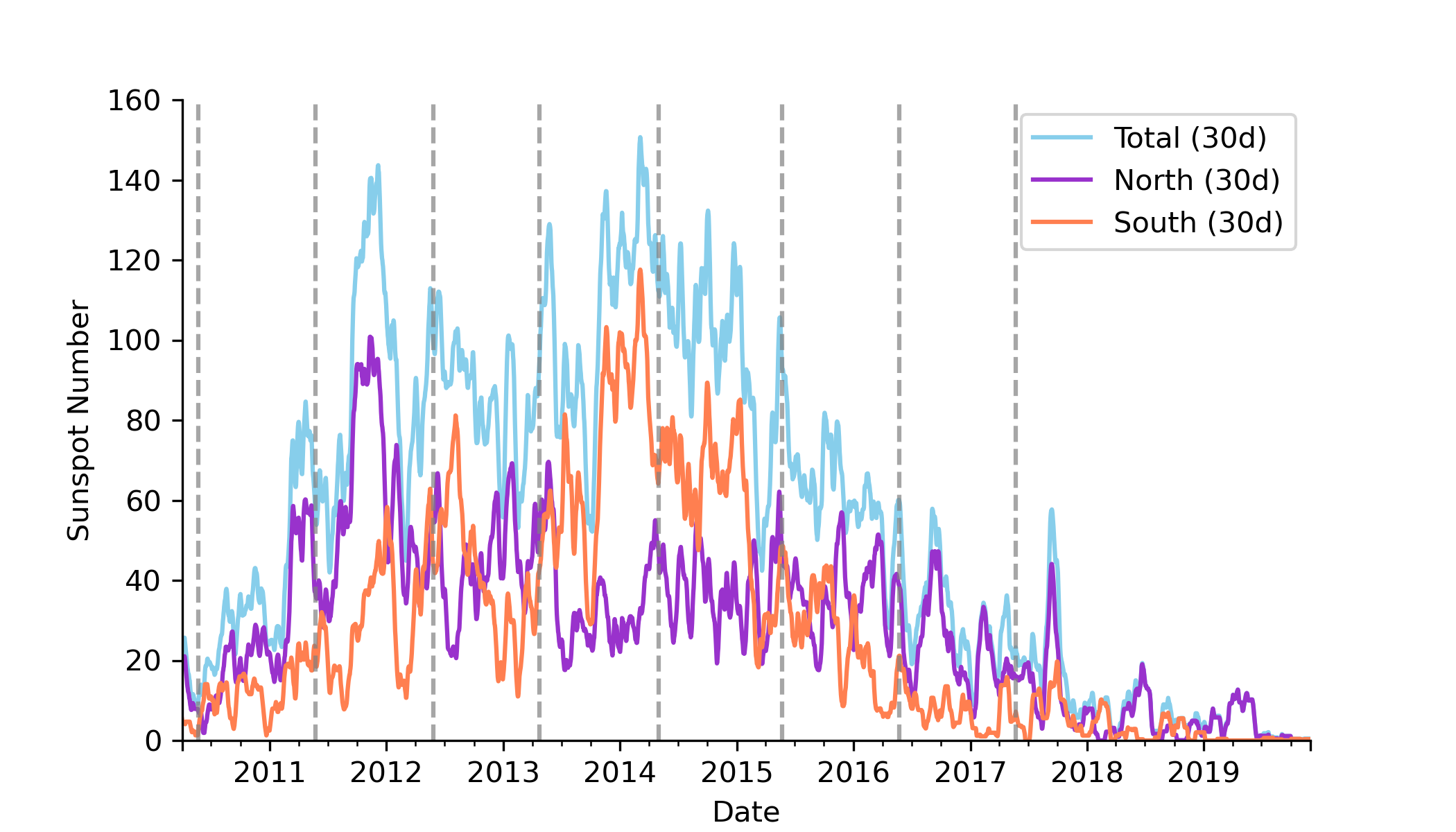}
    \includegraphics[trim=50 0 5 40, clip, scale=0.44]{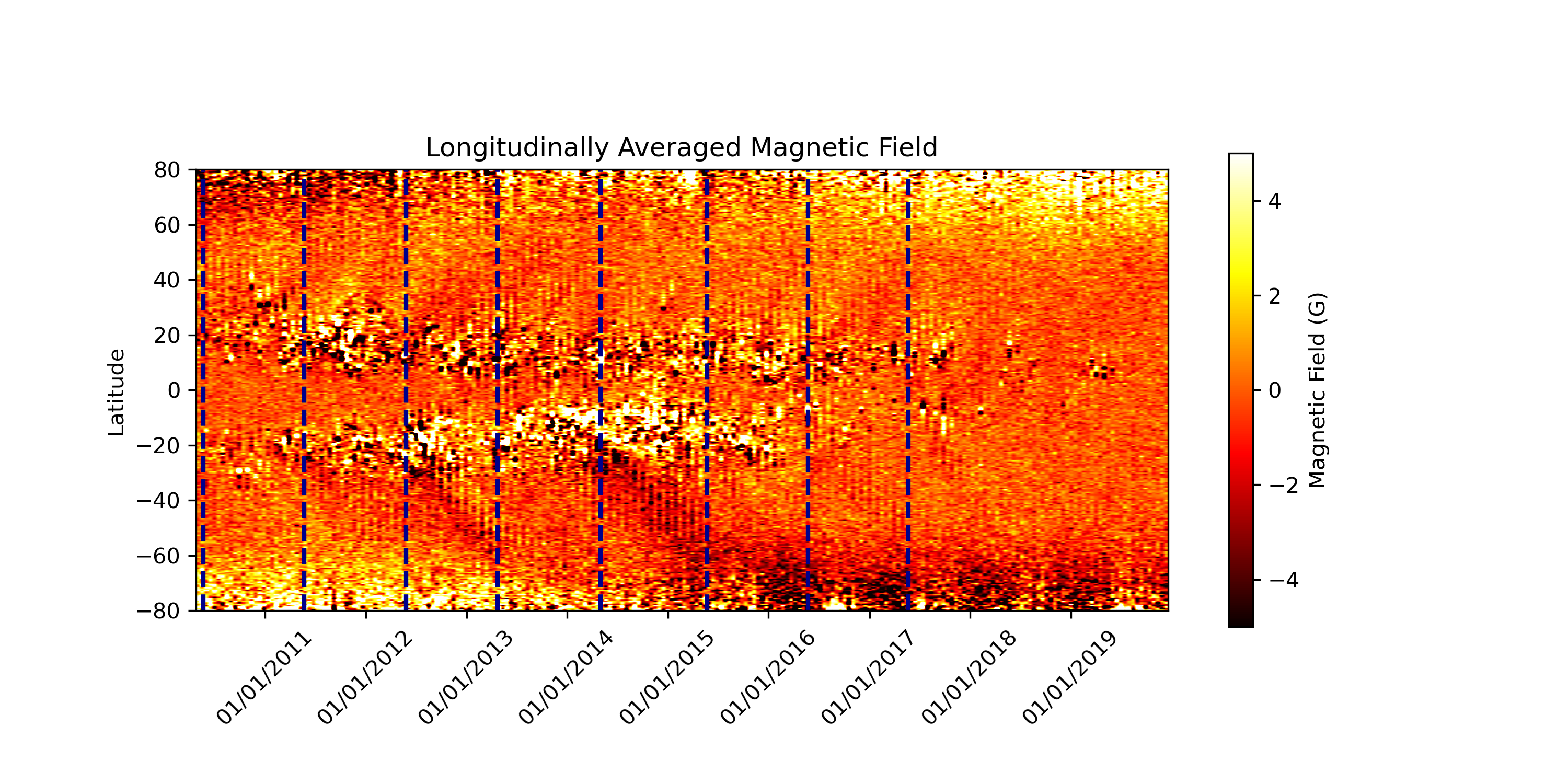}
    \caption{Comparison of solar activity and dates of data sets used. The top panel shows the total (blue), Northern (purple) and Southern (orange) sunspot numbers averaged with a 30 day running mean filter are displayed. The vertical dashed grey lines show the dates corresponding to the coronal hole observations used for this study. The lower panel shows the longitudinally averaged magnetic field. The vertical dashed blues lines show the dates corresponding to the coronal hole observations.}\label{fig:activity}
\end{figure}

The global magnetic field of the Sun is also subject to significant evolution over the solar cycle (e.g., Figure~\ref{fig:activity}), with the polar regions receiving magnetic flux transported from high-latitude decaying active regions, the so-called `rush to the poles' \citep{hathaway_2015}. The polar coronal holes vary in size over the cycle, decreasing towards sunspot maximum and also undergo a reversal of the dominant polarity \citep{Hess_Webber_2014,Illarionov_2020}. This variation in the structure of the magnetic field should have some impact on the wave energy able to reach the corona, as Alfv\'enic wave propagation (e.g., amplification and reflection) are determined by variations in Alfv\'en speed along the magnetic field. Further, the transport of opposite polarity flux to the polar regions may lead to increases in the number and size of reconnection events, which are also known to be a source of waves \citep[e.g.,][]{Takeuchi_2001,Cranmer_2018}.
\medskip

In this work, our goal is to examine whether there are is any evidence for variation in the properties of Alfv\'enic waves in polar coronal holes over the solar cycle and estimate the associated Poynting flux. We use imaging data from the Solar Dynamics Observatory \citep[SDO -][]{PESetal2012} to measure the wave properties, utilising automated wave measurement techniques developed previously. Plasma and magnetic quantities are also estimated via differential emission measure and potential field extrapolations, enabling us to estimate a measure for the Poynting flux from the Alfv\'enic waves.

\section{Observations}\label{sec:obs}

The data used for this study were taken with the Atmospheric Imaging Assembly \citep[AIA -][]{LEMetal2012} onboard the SDO. The wave analysis described in the proceeding section is performed on images from the 171~{\AA} channel. We also use images from the 193~{\AA} channel for context. All data are processed to Level 1.5 using the standard \textit{aia\_prep} routine. The data have a cadence of 12~s and a spatial sampling of $0.6^{\arcsec}$ per pixel.

\begin{figure*}
  
  \begin{adjustbox}{addcode={\begin{minipage}{\width}}{\caption{%
     SDO/AIA images of the coronal holes observed between 2010 and 2013. The top row shows the 171~{\AA} images and the bottom row is from 193~{\AA}. The white lines overplotted in each image show an example
     of the artificial slit (located at 5~Mm above the limb) used for measuring the transverse displacements within the coronal hole.}\label{fig:poles}
      \end{minipage}},rotate=270,center}
      \adjincludegraphics[trim={0cm,0cm,5cm,10cm}, scale=0.90, clip]{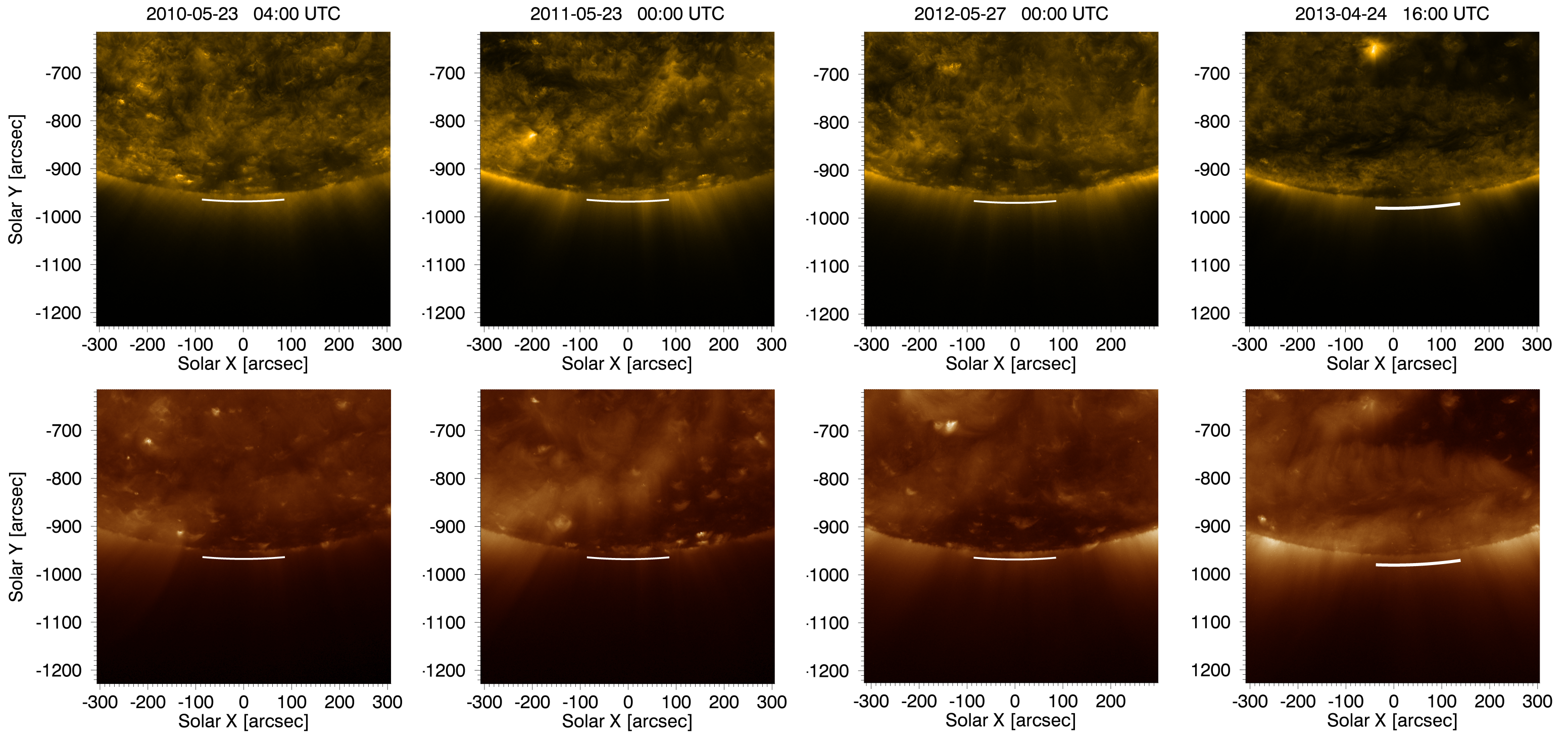}
  \end{adjustbox}

\end{figure*}

\begin{figure*}
      \begin{adjustbox}{addcode={\begin{minipage}{\width}}{\caption{%
      Same as Figure~\ref{fig:poles} but for coronal holes observed between 2014 and 2017.
      }\label{fig:poles2}
      \end{minipage}},rotate=270,center}
      \adjincludegraphics[trim={0cm,0cm,5cm,10cm}, scale=0.90, clip]{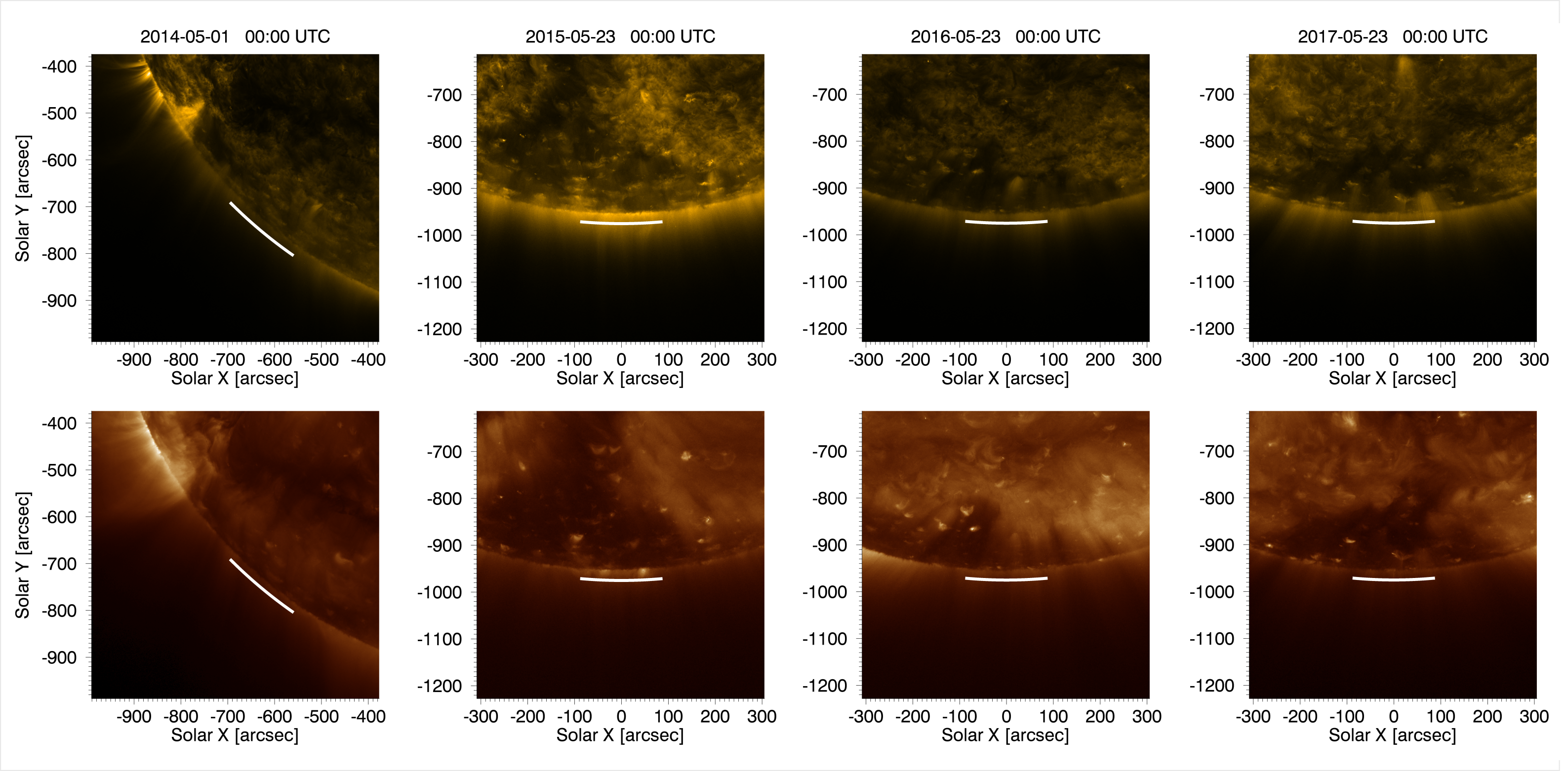}
      \end{adjustbox}
      
\end{figure*}

To examine whether there are is any evidence for variations in the flux of Alfv\'enic waves through the solar cycle, we select 8 data sets between 2010 and 2017 which covers the rise and fall of solar cycle 24. The data sets are chosen to be roughly equally spaced in time, occurring in May (or as close as possible) of each year. We made the decision to limit our analysis to southern polar coronal holes where possible. This was to avoid incorporating any phase differences between the northern and southern polar field reversals \citep[e.g.,][]{Karna_2014}.  The only year that we couldn't find a polar hole close to May was in 2014. As such, we used a coronal hole that occurred at a slightly higher latitude. The selection of coronals holes used is displayed in Figures~\ref{fig:poles} and \ref{fig:poles2}. Each AIA data set for the wave analysis consists of images spanning a 4~hour time period.

For context, we examine the magnetic activity during the different observation dates. The top panel of Figure~\ref{fig:activity} displays the Total, North, and South sunspot numbers (averaged with a 30 day running mean filter) as an indicator of the solar activity. This data was obtained from the Solar Influences Data Analysis Center operated by the Royal Observatory of Belgium\footnote{Data obtained from \url{https://www.sidc.be/SILSO/home}.}. The lower panel shows the longitudinally averaged magnetic field between 2010 and 2019. The averaged magnetic field was obtained using data from the Helioseismic and Magnetic Imager \citep[HMI - ][]{SCHetal2012}. We indicate in each panel of the figure the dates chosen for analysis of the AIA observations. Over the course of the total observation period, the Sun's global magnetic field is changing significantly. The presence and variation in sunspot number is likely to have a minor direct influence on the dynamics of the polar coronal holes. However, the observational dates also span the field reversal at the poles, which occurs sometime between the 2014 and 2015. During this time there is potentially more reconnection due to increased flux cancellation from the greater presence of mixed polarities at the poles. Reconnection is able generate additional waves over granulation or \textit{p}-mode driving and could impact upon the distribution of wave properties in the coronal holes.  

\medskip

We also make use of the 10.7~cm radio flux data which was obtained from the Solar Radio Monitoring Program service operated by the National Research Council and Natural Resources Canada\footnote{Data obtained from \url{https://www.spaceweather.gc.ca/forecast-prevision/solar-solaire/solarflux/sx-5-en.php}}.
\medskip

For differential emission measure analysis, we use SDO/AIA data from the 94~{\AA}, 131~{\AA}, 171~{\AA}, 193~{\AA}, 211~{\AA}, and 335~{\AA} channels. These data are processed differently from the wave observation data. We remove the scattered light using point spread function deconvolution following \cite{grigis_2012}, correct for degradation, rotate and interpolate each channel to the same plate scale, and align the channels. Finally, the intensities for each channel are normalised by the exposure time.

\medskip

The final data product we utilise are GONG zero-point-corrected hourly synoptic maps of the photospheric magnetic field\footnote{Data obtained from \url{https://gong.nso.edu/data/magmap/}}. A list of the maps used in given in Table~\ref{table:gong} with the Carrington rotation number.

\begin{table}
\centering
\caption{Information for GONG data}
\begin{tabular}{cc}
\hline
Carrington & UTC Timestamp   \\
Rotation & \\
\hline
 2097 & 23-05-2010 23:54 \\
 2110 & 23-05-2011 10:14 \\
 2124 & 27-05-2012 11:54 \\
 2136 & 24-04-2013 01:14 \\
 2150 & 01-05-2014 11:04 \\
 2164 & 23-05-2015 11:14 \\
 2177 & 23-05-2016 11:04 \\
 2191 & 23-05-2017 11:14 \\
 \hline
\end{tabular}\label{table:gong}
\end{table}

\section{Analysis}

\subsection{Wave analysis}
We are interested in tracking and measuring the transverse motions of the fine scale plasma structure that is present in coronal holes \citep{MorCun2023}. The measurement of the transverse motions in the AIA data largely follows that discussed in \cite{Weberg_2020}, hence we only briefly review the analysis steps here.

\medskip

\begin{figure}
\centering
    \includegraphics[scale=0.4]{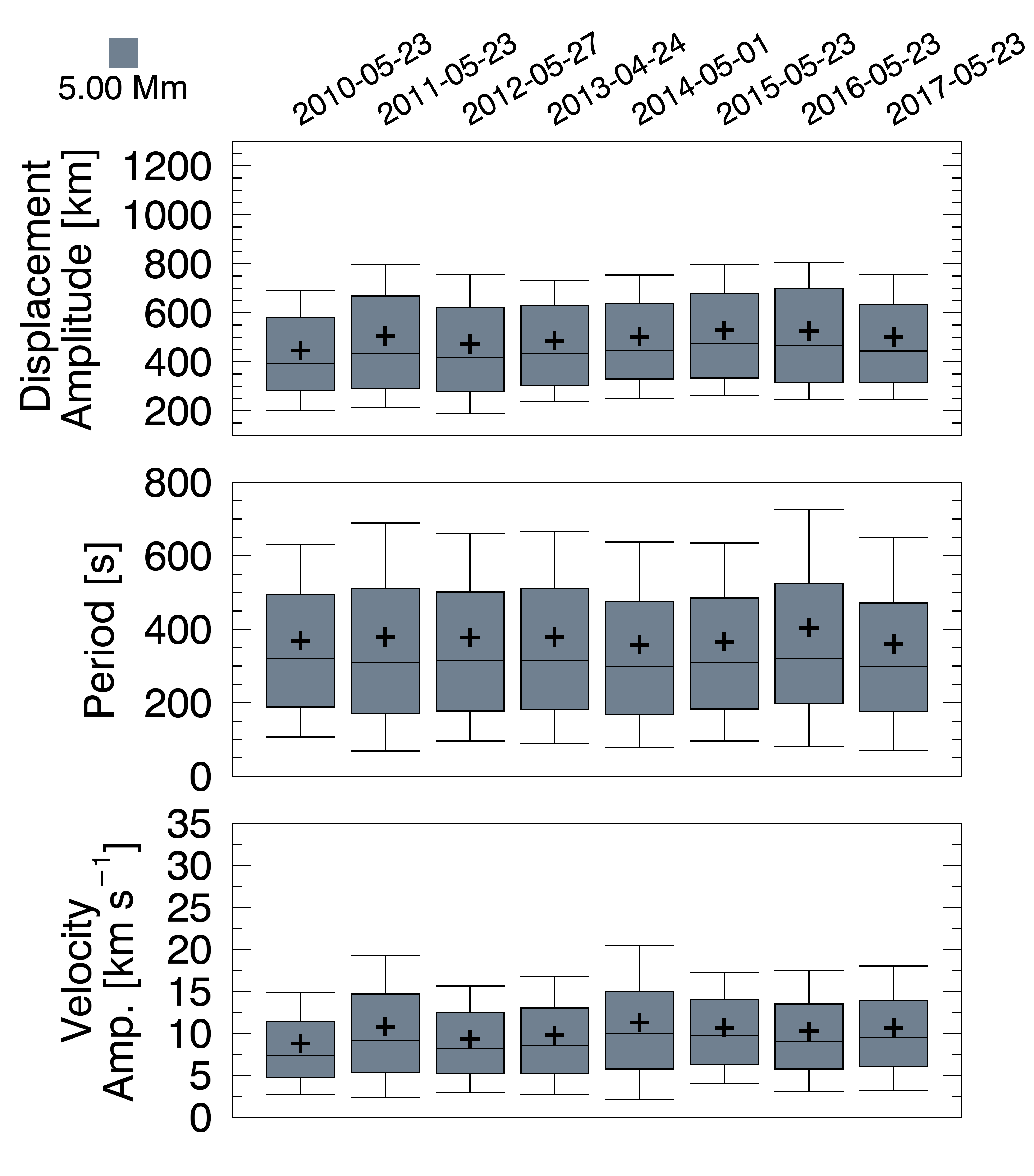}
    \caption{Box plots for wave observations in southern coronal holes taken at 5~Mm above the limb during the different years. The lower and upper edges of the boxes indicate the 1st and 3rd quartiles. The horizontal lines inside the boxes show the median values. The crosses indicate the log-normal means and the whiskers show 1 standard deviation. The distributions are highly skewed, which is why the standard deviation is large.}\label{fig:box5}
\end{figure}

\begin{figure}
\centering
    \includegraphics[scale=0.4]{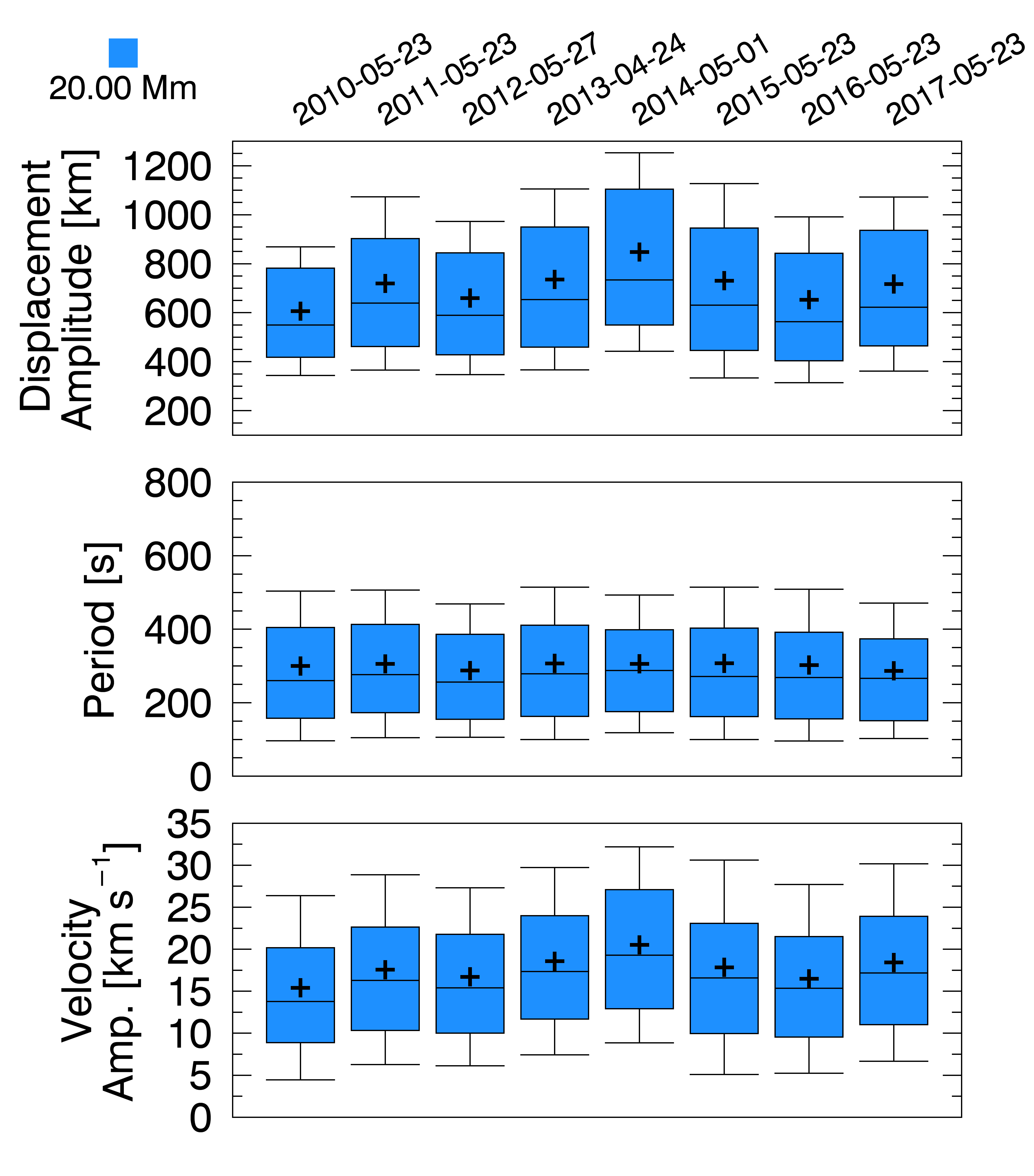}
    \caption{Same as Figure~\ref{fig:box5} but for wave observations in southern coronal holes taken at 20~Mm above the limb during the different years.}\label{fig:box20}
\end{figure}

The following approach is performed for each set of AIA 171~{\AA} data. To isolate the fine-scale plasma structure from the background emission, we apply an unsharp mask filter (6''$\times$6'') to each individual image in the time sequences. We then sample data from seven arc-shaped artificial slits at different heights in the corona, spanning 5-20~Mm (1.007-1.029~$R_\odot$). The slits are separated by 2.5~Mm intervals. Examples of slit locations within the coronal holes for each data set are shown in Figures~\ref{fig:poles} and \ref{fig:poles2}. Sampling the slits over the entire time sequence enables us to create time-distance diagrams at the fixed altitudes.

\medskip

To identify and track the fine-scale structure in the time-distance diagram, we use the NUWT code\footnote{Available at \url{https://github.com/Richardjmorton/auto_nuwt_public}} \citep[a detailed discussion of the code is given in ][]{WEBetal2018}. The code is able to follow the fine-scale features over time and extract individual time-series of the features positions. Each of these time-series identified in the time-distance diagrams is then subject to a Fourier transform and the significant wave components in the time-series are identified. Each slit has between 450-530 wave measurements over the 4 hour period.

\medskip

We repeat the caveat given in \cite{Weberg_2020} here. The measurements do not account for the possible polarisation of the waves relative to the plane of observation. Therefore, the given amplitudes are a lower bound of the true values, likely underestimated by approximately a factor of $\sqrt{2}$.

\subsection{DEM analysis}
In order to examine the plasma properties in the coronal hole, we calculate the Differential Emission Measure (DEM) by inversion of the observed intensities from the six coronal channels of AIA \citep[e.g.,][]{hannah_2012}. The results of the inversion can then provide measures of the temperature and density. In order to obtain the DEM, we use a neural network that has a similar architecture to that of DeepEM \citep{wright_2019} but trained using DEMs generated from the regularised inversion of \cite{hannah_2012}. Full details of the network are provided in \cite{balodhi_2024}. 

From the DEM, a measure of the temperature is given by the first moment of the DEM, i.e.,
\begin{equation}
    \log T_{\mbox{DEM}}=\mbox{DEM}^{-1}\left(\sum_i\mbox{DEM}_i\log T_i\right),
\end{equation}
where the summation is over the number of temperature bins in the DEM inversion (this is also referred to as the `DEM-weighted' temperature). The calculation is performed in $\log T$ space to be consistent with the regularised DEM output. Further, the column emission measure is related to the DEM or density as such:
\begin{equation}
    EM=\int\left(\frac{d\mbox{EM}(T)}{dT} \right)\,dT=\int n_e^2 dz,
\end{equation}
assuming the ion density equals the electron density. Here, $d\mbox{EM}(T)/dT$ is the DEM. The integral in $z$ is along the line of sight. The density of the plasma is then approximately given by:
\begin{equation}\label{eq:den_EM}
    n_e=\sqrt{\frac{EM}{z_{eq}}},
\end{equation}
where $z_{eq}$ is the equivalent column depth of the plasma. Determining the column depth for the coronal holes is non-trivial. We choose to follow the approach outlined in \cite{ASCACT_2001}, which gives an equivalent column depth above the limb as (their Eq.~12):
\begin{equation}\label{eq:z_eff}
    z_{eq}(h)=\int_{-\infty}^{\infty}\exp\left[ -\frac{2}{H_p}(\sqrt{(R_\odot+h)^2+z^2}-r_\odot)\right]dz,
\end{equation}
where $h$ is the height above the limb in solar radii and $R_\odot$ is the solar radius. The pressure scale height, $H_p$, is given by
\begin{equation}
H_p=\frac{k_B T}{\mu m_p g_\odot},
\end{equation}
where $\mu m_p$ is the average ion mass and $g_\odot$ is solar gravitational acceleration.

\medskip

When estimating the wave energy flux for each coronal hole, we will only use the average values of quantities as a function of radial distance. To obtain these, we average the plasma parameters (log temperature and emission measure) for pixels within the coronal holes that lie at fixed radial distances above the limb.

\subsection{Coronal magnetic field estimates}
In addition to the DEM, we also want to obtain insights into the magnetic field strength of the different polar coronal holes. To this end, we utilise potential field source surface extrapolations performed with \textit{pfsspy} \citep{Stansby2020}. The input for the extrapolations are GONG zero-point-corrected hourly synoptic maps of the photospheric magnetic field. To determine the field lines, we select seed points distributed evenly in longitude and latitude over the coronal holes. The field lines are calculated on a 3D grid with 60 points in the (log) radial direction and a source surface at 2.5~$R_\odot$. We are only interested in the average magnetic field strength within the coronal hole. Hence, for each open field line, we interpolate onto a common radial grid and take the average value at each grid location.

\section{Results and discussion}
\subsection{Distributions of wave proprieties}\label{sec:results_perp}
The wave properties we find are broadly similar to previous measurements of the transverse motions in coronal holes \citep{MCIetal2011,THUetal2014,MORetal2015,MORetal2019,Weberg_2020}. Figure~\ref{fig:box5} displays the wave properties (i.e., displacement amplitude, period and velocity amplitude) measured at 5~Mm above the limb for the different coronal holes. The wave properties are shown as box and whisker plots, which enables a comparison of the distributions between years. It can be seen the distributions for all properties appear relatively consistent across the years. For periodicity there is some variation in means, medians and interquartile ranges, although it is only slight (see Table~\ref{tab:period_table}). This indicates that the wave driver is largely consistent over the course of the solar cycle, exciting waves with a continuous range of periods (frequencies). There is more variation in the amplitudes, with mean/median velocity amplitude varying by around 20-30\% over the different years.

In Figure~\ref{fig:box20}, we show the wave properties measured at 20~Mm above the limb. Direct comparison between Figures~\ref{fig:box5} and \ref{fig:box20} shows that the wave amplitudes (displacement and velocity) have increased with height. This is expected due to the dependence of wave amplitude on plasma density and the decrease in density with altitude in coronal holes \citep[see also, e.g.,][]{MORetal2015,Weberg_2020}. The mean value of the period consistently decreases with height, which could be an artefact from the decreasing signal to noise with height, or an indication of the reflection of low frequency waves \citep[also found in][]{Weberg_2020}. We do not investigate this aspect further here.

\begin{figure}
\centering
    \includegraphics[scale=0.32]{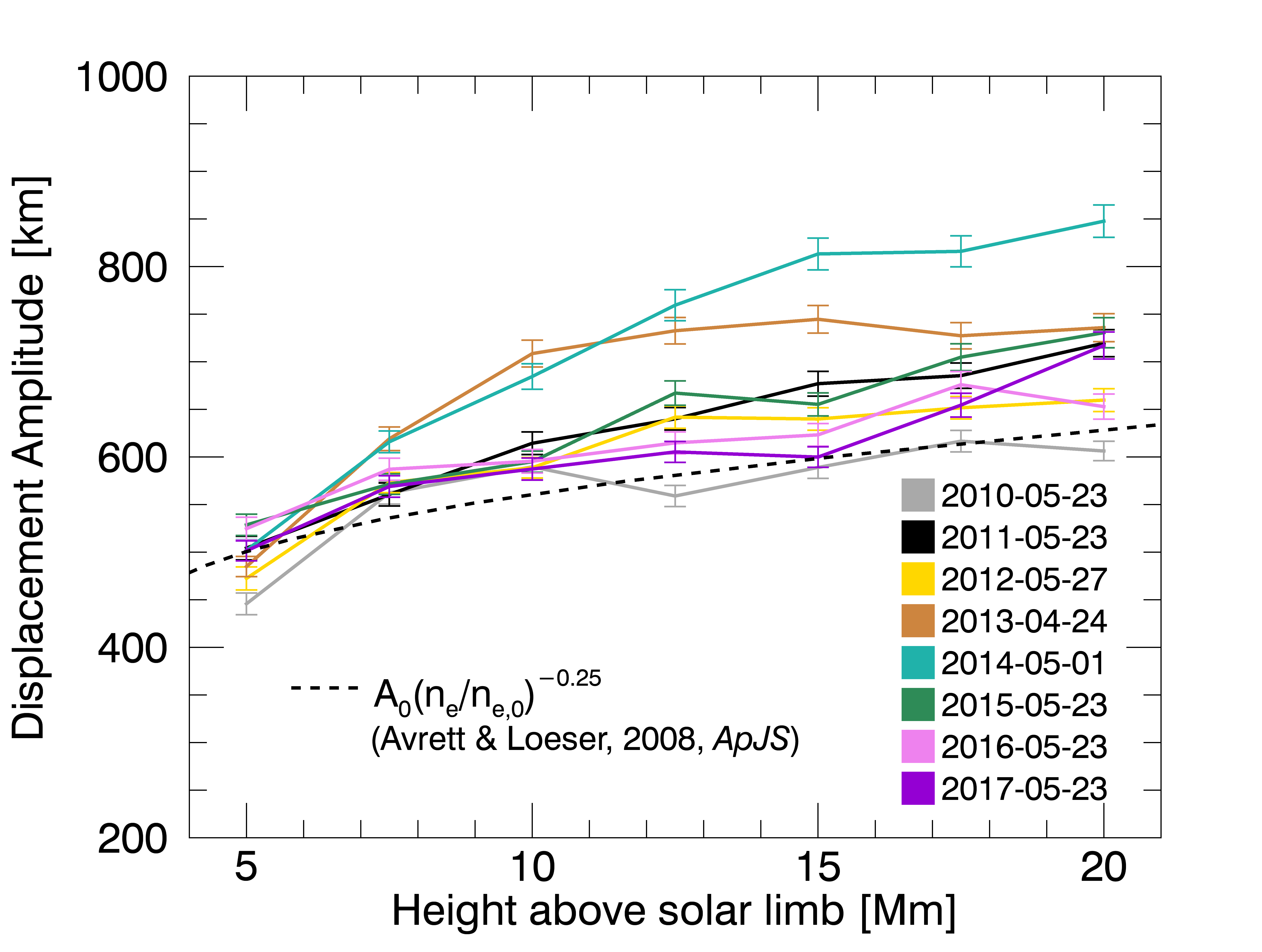}
    \caption{Displacement amplitude as function of height for all coronal holes. The data points and error bars correspond to the mean and standard error, respectively, for the distributions of wave properties at each height. Also shown in the expected variation in amplitude due to variations in density, were we have used the density profile from the semi-empirical model of \cite{Avrett2008}.}\label{fig:disp_vs_h}
\end{figure}

Upon comparison of Figures~\ref{fig:box5} and \ref{fig:box20}, it is also noticeable that there is a much greater variability in the amplitude distributions at 20~Mm above the surface. To further elucidate this, in Figure~\ref{fig:disp_vs_h} we plot the displacement amplitude as a function of height from the different slits. The amplification of the waves with height is clear across all years and broadly follows the change in amplitude expected from decreasing density \citep[comparison provided with density profile from semi-empirical model of ][]{Avrett2008}. From Figures~\ref{fig:box20} and \ref{fig:disp_vs_h} there is no clear evidence to suggest that the observed variation in wave behaviour is influenced by the evolving magnetic conditions in the poles over the solar cycle.

\subsection{Coronal hole properties}
\begin{figure*}
\centering
    \includegraphics[scale=0.85]{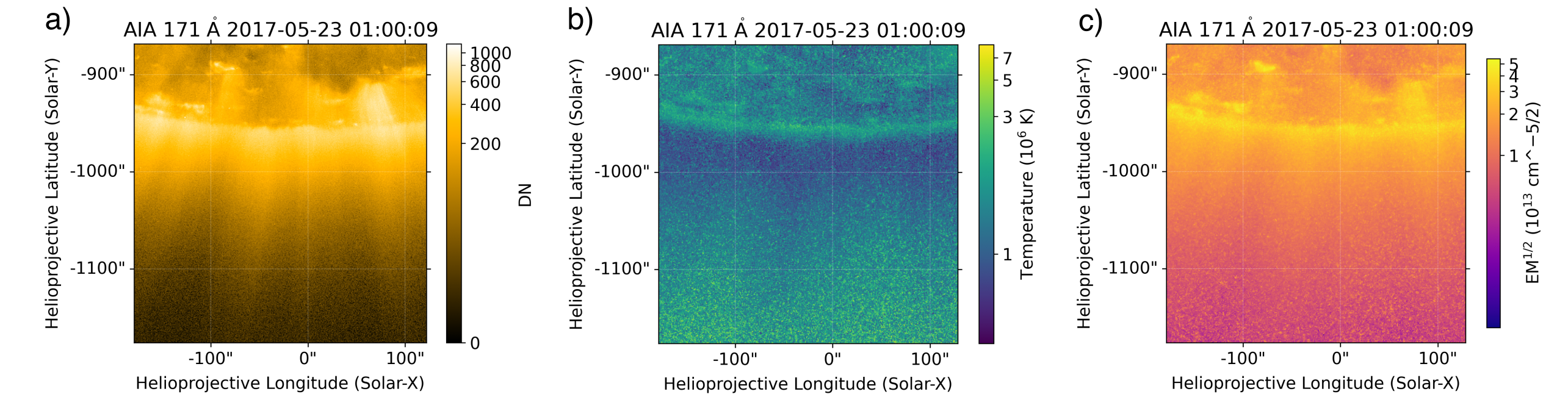}
    \caption{Results of Differential Emission Measure (DEM) analysis. The figure shows the DEM results for the 2017 data set. Panel a) shows the intensity in 171~\AA channel. Panel b) displays the estimated temperature and panel c) is the square root of the total emission measure.}
    \label{fig:ch_props_2017}
\end{figure*}

\begin{figure*}
    \centering
    \includegraphics[scale=1.1]{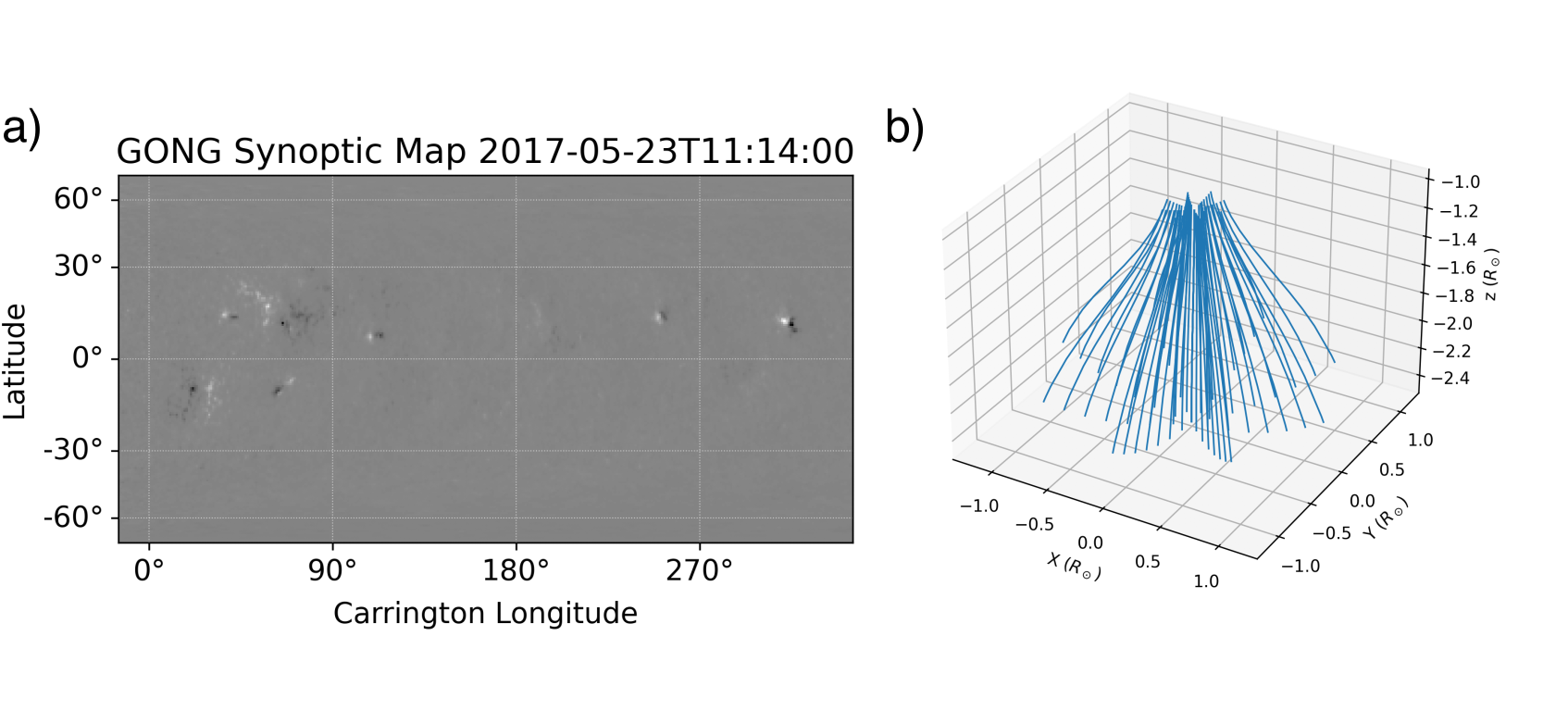}
    \caption{Magnetic field extrapolations. Panel a) shows GONG magnetogram from 2017 during Carrington rotation 2191. Panel b) is the results of the potential field extrapolation of the southern polar coronal hole, which is dominated by open field lines. Note the $Z$ axis is only shown from the solar surface.}
    \label{fig:ch_mag_2017}
\end{figure*}

Figure~\ref{fig:ch_props_2017} shows an example of the results from the DEM analysis, focusing on data from 2017. This year is chosen since 2017 is the lowest quality SDO/AIA data in the sample as it suffers from the greatest amount of instrument degradation (which is increasing over time). Hence it demonstrates the lowest quality inversions. In spite of this, the DEM inversion quality is still high. Temperatures in the coronal hole are between 0.9~MK and 1.2~MK for this year, which is in line with previous results \citep[e.g.,][]{Wilhelm_1998,Fludra_1999,Young_1999,Landi_2008,CRA2009,Hahn_2010}. The square root of the emission measure is shown as a proxy for electron density. The emission measure shows an expected decreases with height above the limb. Based upon the solution to Eq.~\ref{eq:z_eff}, one would expect the density to fall off only marginally less quickly. The magnetic structure for the 2017 coronal hole is also determined from a potential field extrapolation. Figure~\ref{fig:ch_mag_2017} displays the GONG synoptic map used (panel a) and the results of the extrapolation (panel b). Only the field lines originating from the southern polar region are shown.

\begin{figure*}
\centering
    \includegraphics[scale=0.9]{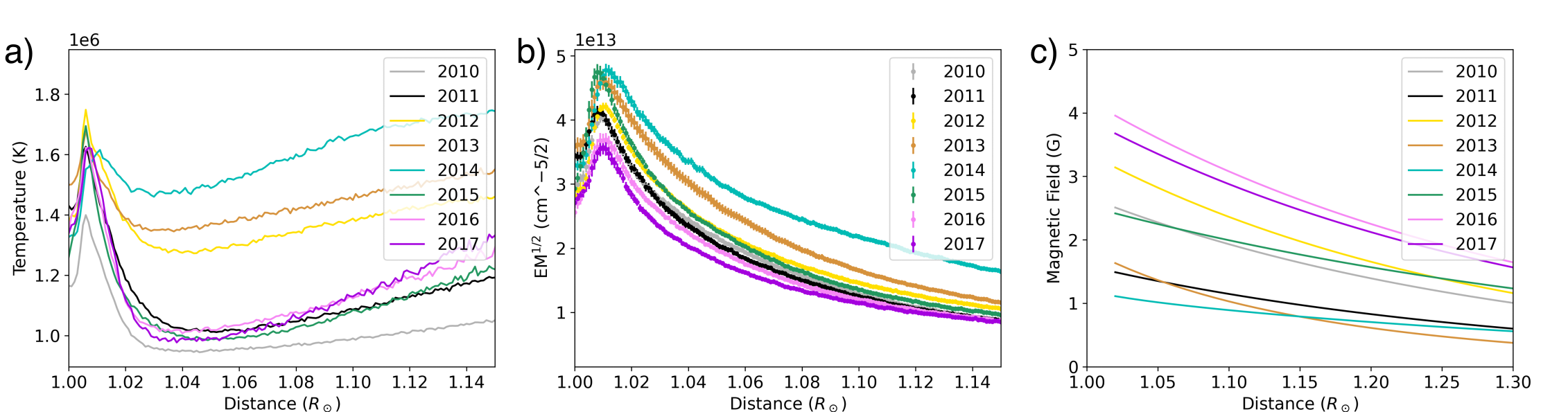}
    \caption{Estimated magnetic and plasma properties for the coronal holes. Panels a) and b) show the average temperature and square root of the emissions measure, respectively, across the coronal hole estimated from the DEM analysis. Panel c) shows the estimated magnetic field averaged over the coronal hole from the potential field extrapolations.}
    \label{fig:ch_props}
\end{figure*}

The averaged radial profiles of square root of emission measure, temperature and magnetic field strength are shown in Figure~\ref{fig:ch_props} for all years. In general, the properties across the coronal holes are broadly comparable and are in line with previous measurements given for coronal holes \citep[e.g.,][]{Wilhelm_1998,Fludra_1999,Young_1999,Landi_2008,CRA2009,Hahn_2010}. The emission measure shows an initial increase up to 1.01~R$_\odot$, which is due to an effective doubling of the plasma column depth after moving above the limb and chromosphere. After this, it decrease in line with expectations from a decreasing density. The temperature also shows an initial increase near the limb which is unexpected. Many previous analysis of coronal holes tend to avoid reporting measurements across the limb, however a similar increase has been reported before in \cite{Fludra_1999}. The temperature increase has also been confirmed to be present in Hinode EIS data (Private communication with P. Young), so we believe it is genuine. We will not follow up this here though. 

The coronal holes from 2013 and 2014 appear to be hotter, denser and with weaker magnetic fields when compared to the others. The reason for this could be because the 2013 coronal hole is smaller and less well defined than the other coronal holes, and the 2014 coronal hole is the only non-polar hole. The evolution of the waves with height from these two coronal holes can be seen to stand apart from the wave evolution in other coronal holes (Figure~\ref{fig:disp_vs_h}).

\subsection{Poynting fluxes}
An important aspect is to determine whether there is any variation in wave energy flux through the coronal holes over the solar cycle. There are a number of issues associated with calculating the wave energy flux from imaging observations and it is worth engaging in a discussion before continuing. 

\medskip
In general, the Poynting flux is given by
\begin{equation}
    \vec{S} =\frac{1}{2\mu_0}(\vec{v}\times\vec{B})\times\vec{B}. 
 \end{equation} 
{Hence there are clearly contributions from flows and waves to the Poynting flux. In coronal holes, there is the persistent outflow that forms the solar wind but at the  heights we make the wave measurements, we do not have to consider its contribution to the Poynting flux. The separate contributions from flows and waves can be obtained by writing the vectors as the background quantity plus a perturbation, i.e., $\vec{v}=\vec{v}_0+\vec{v}_1$ and $\vec{B}=\vec{B}_0+\vec{B}_1$, where the subscript 0 indicates a background quantity and 1 a perturbation. By linearising the equation for Poynting flux (i.e., neglecting products of perturbations), there are terms proportional to $v_1 B_0^2$ or $v_0 B_1^2$. The term consisting of only background components is zero as the flow is along the magnetic field.}

The relationship between velocity and magnetic field perturbations for Alfv\'enic waves \citep[both Alfv\'en and kink modes – see][]{SPR1982} is $v_1 B_0/c_{ph}=B_1$, where $c_{ph}$ is the phase speed. This means that $v_0 B_1^2=v_0 v_1^2 B_0^2/c_{ph}^2$. Hence, the contributions of the flow terms to the Poynting flux scale as $v_0/c_ph^2$.  The outflow speed at $<1.2 R_\odot$ is predicted to be on the order of 10~km/s \citep[e.g.,][]{Pinto_2009,Shoda_2018}, while the propagation speed is 300-500~km/s \citep{MCIetal2011,MORetal2015}. Hence, the contributions to the Poynting flux from the solar wind flow at the current heights are negligible compared to the contributions from Alfv\'en waves.

The Alfv\'enic wave energy flux can be calculated using the time-averaged, area integrated vertical Poynting flux, $\langle S_z\rangle$, which is given by \citep[e.g.][]{GOOetal2013}:
\begin{equation}\label{eq:poynting}
    \langle S_z\rangle=\frac{1}{2}(\rho_i+\rho_e)v^2R^2c_k,
\end{equation}
which is based upon approximating the density enhancements in the solar corona as a magnetic cylinder (of radius $R$) with an internal, $\rho_i$, and external density, $\rho_e$. Here $c_k$ is the kink speed, given by 
$$
c_k^2=\frac{2}{\rho_i+\rho_e}\frac{B_0^2}{\mu_0}.
$$
The comparative formula for the vertical Poynting flux of the bulk Alfven wave is \citep{GOOetal2013}:
\begin{equation}\label{eq:poynt_alf}
    \langle S_z\rangle=\frac{1}{2}\rho_iv^2\alpha R^2v_{Ai},
\end{equation}
when integrating over a cylinder of radius $\alpha R$ and $v_{Ai}=B/\sqrt{\mu_0\rho_i}$. \textbf{We note here that $c_k$ and $v_{Ai}$ are the group speeds of the Alfv\'enic waves.}

 \cite{GOOetal2013} suggested that using Eq.~\ref{eq:poynt_alf} instead of Eq.~\ref{eq:poynting} leads to an overestimation of the energy of the kink wave by:
$$
\frac{\rho_i+(\alpha^2-1)\rho_e}{\rho_i+\rho_e}\leq F \leq\alpha^2\frac{\rho_i}{\rho_i+\rho_e}
$$
where $F$ is a multiplicative constant and $\alpha$ is a measure for the inverse filling factor of discrete overdense flux tubes. However, in coronal holes, the overdensity of the flux tubes is fractional. The intensity of the overdense structures in the coronal holes in SDO images is only around 2-3\% of the background plasma \citep{MorCun2023}, suggesting that $\rho_i/\rho_e\approx 1$ is a reasonable approximation. This is supported by the lack of observable damping for the Alfv\'enic waves in coronal holes \citep[e.g.,][]{MORetal2015, Weberg_2020} as the damping rate is dependent upon the density difference between internal and external plasma \citep[e.g.,][]{TERetal2010c,VERTHetal2010}. In this case, the above expression reduces to $F\approx{\alpha^2}/{2}$. This also implies the kink speed reduces to $c_k\approx B/\sqrt{\mu_0\rho}$.

The filling factor of the flux tubes is also unknown in the corona. However, recent radiative MHD simulations of the corona suggest that the presence of discrete flux tubes may be an artefact of line-of-sight integration through an optically thin medium \citep{Malnushenko2022,Kohutova_2024}, with a hierarchy of turbulent-like features present across various scales. The hierarchy of scales is supported by observations which show power-law behaviour for the spatial structure of the intensity in images (down to the resolution limit), both in the inner corona \citep{MorCun2023} and outer corona \citep{DeForest_2018}. Hence the filling factor may be close to one. This means that the difference between using Eq.~\ref{eq:poynting} and Eq.~\ref{eq:poynt_alf} is likely minimal for coronal holes.

Using the approximation that $\rho_i\approx\rho_e$, the time-averaged Poynting flux for the kink mode is
\begin{equation}\label{eq:poynting_2}
    \langle S_z\rangle\approx\sqrt{\frac{\rho}{\mu_0}}Bv^2,
\end{equation}
where $\rho$ is the mass density, $B$ is the magnetic field strength and $v$ is the velocity amplitude of the wave. Note this not area-integrated like Eq.~\ref{eq:poynting}. Eq.~\ref{eq:poynting_2} is, essentially, equivalent to the time-averaged Poynting flux for the bulk Alfv\'en and is the one often used in observational estimates of wave energy. We choose to use Eq.~\ref{eq:poynting_2} for our calculations of Poynting flux given we believe it provides a reasonable estimate for the energy flux. It is also comparable to the expression used in the estimation vertical Poynting flux from the numerical simulations of \cite{Hazra_2021} and \cite{Huang_2023,Huang2024}.

\medskip

\begin{figure}
\centering
    \includegraphics[trim= 22 35 100 170 ,clip, scale=0.8]{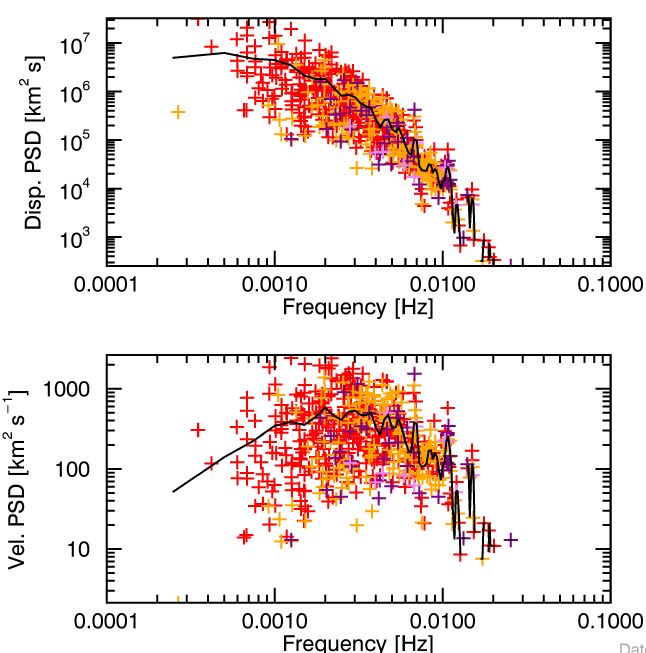}
    \caption{Example power spectral densities for displacement amplitude (top) and velocity amplitude (bottom) for the wave measurements from the 2015 data set. The crosses are the individual wave measurements and the black is the result of a non-parametric regression to the data points.}
    \label{fig:ch_2015_PSD}
\end{figure}

There is also an unresolved issue around picking values of velocity amplitude to use within Eq.~\ref{eq:poynting_2}. Previous studies have taken the average value of velocity amplitude from a distribution of measured values. However, it is not obvious this is the correct thing to do. As known from analysis of Doppler velocity time-series, fluctuations are continually present in the corona across the observable frequency range \citep[e.g.,][]{TOMetal2007,MORetal2015,MORetal2019}. The analysis of imaging observations does not measure this continual level of fluctuation, but only the wave motions that are present on structures that stand out from the background. Upon choosing the average velocity amplitude, this effectively selects the energy flux at a single frequency, as wave amplitude and frequency are correlated \citep[e.g., see joint distributions of wave amplitude and frequency in][]{MORetal2019,Weberg_2020}. 

Ideally, to estimate a total wave energy flux, one should create a power spectrum from wave measurements in the imaging observations and the total energy across all frequencies should be calculated. An example of displacement and velocity power spectral densities \citep[calculated following][]{MORetal2019} for the waves in the 2015 coronal hole are shown in Figure~\ref{fig:ch_2015_PSD}. However, these power spectra use wave measurements taken across an extended region of the coronal holes (associated with the artificial slit) rather than a single location and combines wave measurements that occur co-temporally (at different locations along the slit). This introduces a challenge in trying to arrive at a scaling factor to account for this. We do not tackle this challenge here, but make the assumption that the variations in wave energy flux calculated from the mean value of velocity amplitude reflect the variation in total wave energy flux. We believe this approach is valid because the distributions of period remain largely unchanged over the solar cycle (various statistical estimators for the distributions are given in Table~\ref{tab:period_table}), while the amplitudes of the waves vary. Given the bivariate relationship between period (frequency) and amplitude, an increase in mean/median wave amplitude implies the amplitude increases across all frequencies. This behaviour is also seen in the Doppler velocity powers spectral densities obtained with CoMP \citep[see Figure~4 in][]{MORetal2019}.

\begin{table}[]
\caption{Period statistics. Values given in seconds. Q1 and Q3 are the lower and upper quartiles.}
\label{tab:period_table}
\begin{tabular}{lllll}
Year & Q1 & Median & Mean & Q3 \\
\hline
 2010 & 157 & 258 & 300 & 403  \\
 2011 & 171 & 273  & 306  & 414  \\
 2012 &  154 & 256  & 286 & 386  \\
 2013 & 162 & 278 & 308  & 410 \\
 2014 &  175 & 288  & 306  & 399  \\
 2015 &  162& 271  & 307  & 403  \\
 2016 &  155& 266  & 302  & 391 \\
 2017 &  151 & 265  & 289  & 374 
\end{tabular}
\end{table}

\medskip

We calculate the vertical Poynting flux at a height of 20~Mm (1.03~$R_\odot$) above the limb for all data sets. Using the values from this altitude avoids the uncertainty associated with the temperature increase from the DEMs that seems to occur near the limb.  In order to estimate the electron density from the Emission Measure via Eq.~\ref{eq:den_EM}, we need to estimate the column depth, $z_{eq}$. Eq.~\ref{eq:z_eff} requires us to provide a temperature for the plasma, hence we take the value estimated at 20~Mm from the DEMs. However, we acknowledge this is a non-unique measure of the temperature of a multi-thermal plasma. The values of the equivalent column depths are between $1.4\times10^{10}<z_{eq}<2.3\times10^{10}$~cm. These values are comparable to those obtained by \cite{Young_1999} using a complementary method. The estimated values for the electron number density are then around $2-3\times10^8$~cm$^{-3}$, which are again in-line with previous estimates. The mass density is calculated as $\rho=\mu m_p n_e$. The magnetic field values are taken from the potential field extrapolations at the same altitude. 

We also aim to provide an uncertainty on the wave flux measurements. The uncertainties are calculated using the standard errors of the average values for wave and plasma quantities. The plasma temperature and density uncertainties are naturally small due to the averaging over the coronal hole. These estimates do not include any bias that might arise from systematic issues related to scattered light, flux correction (from degradation), or response functions. From the potential field extrapolation the only factors we can control are the the placement of seed points. Hence, we vary the location of the seed points to estimate uncertainties and examine the impact on the average magnetic field value. For most dates we find that the average magnetic field fluctuates by $\delta B\sim0.1$~G except for 2014, where the variability is larger. As with density and temperature values, this doesn't capture any bias in using the potential field model.

\medskip

\begin{figure}
\centering
    \includegraphics[scale=0.5]{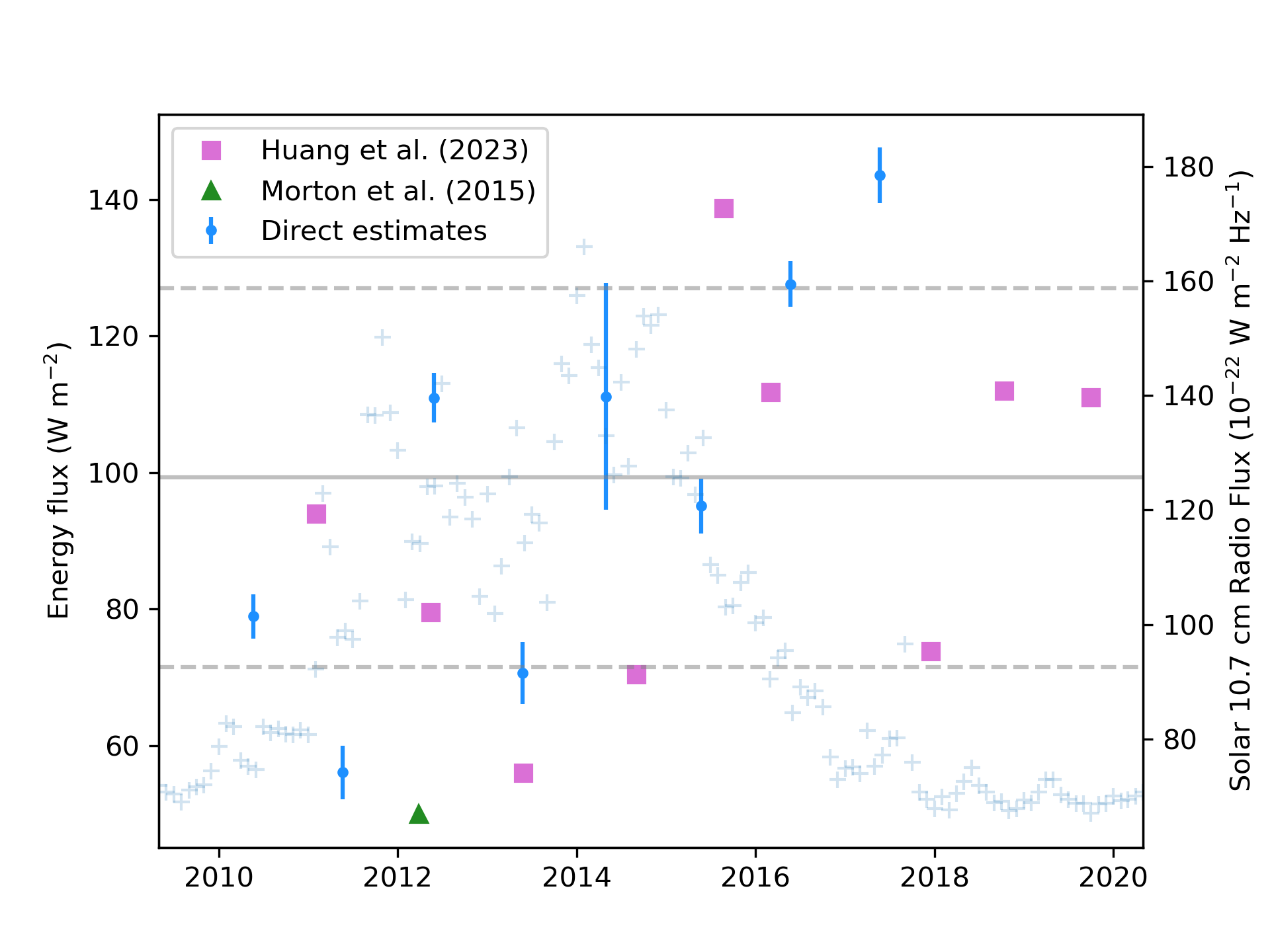}
    \caption{A measure of the Poynting flux in the coronal holes calculated at 20~Mm above the limb. Note this value is based on the average wave amplitude, so is only a measure of Poynting flux as opposed to the true Poynting flux contributed by the waves (see text for further discussion). The shown error bars are the standard error for the estimates. The shown uncertainty does not include bias that may arise during estimates of the magnetic and plasma quantities. The solid horizontal line is the mean wave energy and the dashed lines indicate $\pm1$ standard deviation. Also shown are the wave energy flux estimates from \cite{Huang_2023} (squares - reduced from true value by a factor of 0.2, see text) and the 10.7cm solar radio flux (light blue crosses).}
    \label{fig:ch_wave_flux}
\end{figure}

The measure of vertical Poynting flux for the waves in the different coronal holes is shown in Figure~\ref{fig:ch_wave_flux}. The estimates reveal that the energy flux over the solar cycle is relatively constant, with an average value of $\bar{\langle S\rangle}=99$~W~m$^{-2}$ with a standard deviation of $\sigma_S=28$~W~m$^{-2}$. This represents a 30\% fluctuation in vertical Poynting flux between the different coronal holes. We also show the estimated value of Poynting flux from \cite{MORetal2015}. These values are considerably smaller than the values reported in \cite{Huang_2023} and  \cite{Huang2024}, who estimated the vertical Poynting flux ($\bar{\langle S\rangle}=470$~W~m$^{-2}$, $\sigma_S=130$~W~m$^{-2}$ and $\bar{\langle S\rangle}=520$~W~m$^{-2}$, $\sigma_S=140$~W~m$^{-2}$, respectively\footnote{We have corrected these values from those presented in \cite{Huang2024} by the erroneous factor of 10.}) required for the wind generated by the AWSoM model (via wave turbulence) to best match observed velocities and densities of the solar wind across the solar cycle. We note that their vertical Poynting flux also varies by around 30\%.

The majority of the waves measured here have a median periodicity of around 270~s, hence are related to the enhanced coronal power found in CoMP data \citep[see results from][]{MORetal2019}. The enhanced power is thought to arise from mode conversion of \textit{p}-modes \citep{HANCAL2012,KHOCAL2012,MORetal2019}. Hence, one might expect the observed solar cycle dependence on the \textit{p}-mode amplitudes would impact upon the coronal wave energy flux. The \textit{p}-mode amplitudes vary by $\pm17\%$ over the cycle \citep{Kiefer2018} and are known to vary in anti-phase with the 10.7~cm radio flux. In Figure~\ref{fig:ch_wave_flux} we show the wave energy flux estimates obtained here along with the 10.7cm flux. However, the wave fluxes do not visually show any correlation with the 10.7~cm flux (any numerical estimates of correlation would be highly uncertain). We also show the results from \cite{Huang_2023, Huang2024}, who use different dates for their analysis. The Poynting flux estimated from the observations here are likely not directly comparable to the model results from \cite{Huang_2023, Huang2024} as we believe we are underestimating the Alfv\'enic wave energy for several reasons. As mentioned above, the kink wave energy is effectively estimated for a single frequency rather than integrating the energy across frequencies. Further to this, Alfv\'enic wave energy is also likely also present in the corona in other modes (e.g., torsional and fluting modes), although these are yet to be found. Hence in order to contrast the vertical Poynting flux in the simulations to our results, we multiply the results from \cite{Huang_2023, Huang2024} by a constant such that the mean of the simulation results matches the observational value. This assume the two measures of Poynting flux are related by a multiplicative factor. From Figure~\ref{fig:ch_wave_flux}, there is no evident pattern or trend in either the individual data sets or combined, potentially suggesting that the Poynting flux due to waves is independent of solar cycle.

This could be for a number of reasons, one of which is that there is currently too few coronal wave flux estimates available and any relationship is masked by the scatter. The scatter could arise from measurement uncertainty but also could be from variations in coronal hole properties. The plasma and magnetic conditions in the lower solar atmosphere may also have a bigger impact on coronal wave energy flux than the variation in \textit{p}-mode amplitudes.

Alternatively, the \textit{p}-mode amplitudes at the polar regions could be less affected by the global changes in the magnetic cycle. The \textit{p}-mode amplitudes are thought to be influenced by the turbulent convection in shallow layers \citep[e.g.,][]{balmforth_1992,kumar_1994}, hence only the near-surface magnetic activity is likely to influence the amplitude. The studies of mode variation use GONG data and undertake Sun-as-a-star analysis of the \textit{p}-modes, hence the results are likely strongly influenced by the dynamics in the magnetic activity belt. The 10.7~cm flux is also predominantly associated with chromospheric and coronal emission from strongly magnetised regions \citep{Tapping_2013}, hence the correlation between the radio emission and the \textit{p}-mode amplitudes could indicate the tendency for measured mode amplitudes to reflect conditions in the activity belt. In this case, the observed variation would then most likely be dominated by the differences in plasma and magnetic conditions in the lower solar atmosphere of each coronal hole.  

\section{Conclusion}\label{sec:discussion}

Alf\'venic waves are a leading candidate to explain the heating and acceleration of the solar wind, and form the key energy source for many global MHD models which are used to predict solar wind conditions \citep[e.g.,][]{van_der_Holst_2014,Hazra_2021,Shoda_2023}. Many of these models have free parameters, some of which are related to the nature of the Alfv\'enic waves. One aspect that is currently receiving attention is the amount of wave Poynting flux that should be included in the model \citep[e.g.,][]{Hazra_2021,Huang_2023}, which is essentially controlled by the value of Alfv\'enic wave amplitude (as the magnetic field constrained from photospheric magnetograms). In order to provide some guidance on this issue from an observational perspective,
we have presented measurements of coronal Alfv\'enic waves in polar coronal holes from across Solar Cycle 24. The Alfv\'enic waves are present across a broad range of frequencies and the associated distribution varies only minimally. This indicates that there is a consistent driver supplying a continuos energy flux. However, there is a more substantial variation in the waves amplitudes, which show different distributions at different time of the solar cycle.  Upon estimating the magnetic field and density with the coronal holes, we have also been able to provide an estimate for a measure of wave energy flux. The value for the Poynting flux supplied by the waves has an average value of $\bar{\langle S\rangle}=99$~W~m$^2$ with a standard deviation of $\sigma_S=28$~W~m$^2$. Again, we note these values are likely not the total wave energy flux through the coronal holes, but only a measure based on the average wave amplitude. While wave amplitudes and Poynting flux vary over the solar cycle, there is no apparent correlation to magnetic activity (when compared to the 10.7~cm flux). Hence, we suggest the observed variation is likely due to differences between the individual coronal holes.

\textbf{Finally, we note that the wave energy flux estimated here is for time scales greater than $\sim$100~s. We believe these time scales are indicative of the energy containing scales associated with turbulence. There is no clear evidence of fully developed turbulence in the low corona, or even the distinct associated regimes (i.e., energy, inertial dissipative), to confirm this suggestion. However, recent measurements from Parker Solar Probe near the Alfv\'en surface ($>10 R_\odot$) suggest that time scales above 2 minutes are energy containing \citep{Huang__2024}. Given that the associated break point between energy-containing and inertial ranges transitions to longer time-periods with increasing distance from the Sun \citep[e.g.,][]{Chenetal2020,2023ApJ...950..154D}, one might expect that closer to the Sun the break point should be present at shorter time-scales than 2 minutes. Hence, we believe the wave energy estimates provided here provide useful constraints for energy input into Alfv\'enic wave turbulence models.} 

\section{Acknowledgements}

RJM is supported by a UKRI Future Leader Fellowship (RiPSAW—MR/T019891/1). Data used has been provided courtesy of NASA/SDO and the AIA, EVE, and HMI science teams. It is freely available at \url{http://jsoc.stanford.edu/}. For the purpose of open access, the author(s) has applied a Creative Commons Attribution (CC BY) licence to any Author Accepted Manuscript version arising.

This work utilizes GONG data obtained by the NSO Integrated Synoptic Program, managed by the National Solar Observatory, which is operated by the Association of Universities for Research in Astronomy (AURA), Inc. under a cooperative agreement with the National Science Foundation and with contribution from the National Oceanic and Atmospheric Administration. The GONG network of instruments is hosted by the Big Bear Solar Observatory, High Altitude Observatory, Learmonth Solar Observatory, Udaipur Solar Observatory, Instituto de Astrofísica de Canarias, and Cerro Tololo Interamerican Observatory

\section{Facilities}
{SDO (AIA)}

\section{Software}
{Data analysis has been undertaken with the help of NumPy \citep{Numpy}, 
matplotlib \citep{Matplotlib}, IPython \citep{IPython}, Sunpy \citep{sunpy}, Astropy \citep{astropy}, SciPy \citep{Scipy}.}

\bibliographystyle{aasjournal}

\end{document}